\documentclass[twocolumn,letterpaper]{aastex6}

\usepackage{natbib}
\usepackage{amsmath}
\usepackage{txfonts}
\usepackage{gensymb}
\usepackage{mathrsfs}
\usepackage{epstopdf}
\usepackage{graphicx,color}

\newcommand{\rp}{r_{\rm p}}

\newcommand{\rs}{r_{\rm s}}

\newcommand{\Mp}{M_{\rm p}}

\newcommand{\eqnref}[1]{Equation \ref{#1}}
\newcommand{\secref}[1]{\S\ref{#1}}
\newcommand{\figref}[1]{Figure \ref{#1}}
\newcommand{\tabref}[1]{Table \ref{#1}}

\begin{document}

\title{Inner Super-Earths, Outer Gas Giants: \\How Pebble Isolation and Migration Feedback Keep Jupiters Cold}

\author{Jeffrey Fung\altaffilmark{1,3}, Eve J. Lee\altaffilmark{2}}
\altaffiltext{1}{Department of Astronomy, University of California at Berkeley, Campbell Hall, Berkeley, CA 94720-3411, USA}
\altaffiltext{2}{TAPIR, Walter Burke Institute for Theoretical Physics, Mail Code 350-17, Caltech, Pasadena, CA 91125, USA}
\altaffiltext{3}{NASA Sagan Fellow}

\email{email: jeffrey.fung@berkeley.edu, evelee@caltech.edu}

\begin{abstract}
The majority of gas giants (planets of masses $\gtrsim10^2 M_\oplus$) are
found to reside at distances beyond $\sim1$ au from their host stars. Within 1 au, the planetary population is dominated by super-Earths of $2-20 M_\oplus$.
We show that this dichotomy between inner super-Earths and outer gas giants can be naturally explained should they form in nearly inviscid disks.
In laminar disks, a planet can more easily repel disk gas away from its orbit.
The feedback torque from the pile-up of gas inside the planet's orbit slows down and eventually halts migration. A pressure bump outside the planet's orbit traps pebbles and solids, starving the core.
Gas giants are born cold and stay cold: more massive cores are preferentially formed at larger distances, and they barely migrate under disk feedback.
We demonstrate this using 2D hydrodynamical simulations of disk-planet interaction lasting up to $10^5$ years: we track planet migration and pebble accretion until both come to an end by disk feedback.
Whether cores undergo runaway gas accretion to become gas giants or not is determined by computing 1D gas accretion models.
Our simulations show that in an inviscid minimum mass solar nebula, gas giants do not form inside $\sim$0.5 au, nor can they migrate there while the disk is present.
We also explore the dependence on disk mass, and find that gas giants form further out in less massive disks.
\end{abstract}



\section{Introduction}
\label{sec:intro}

Rocky, Earth-sized planets and gaseous, Jupiter-sized planets are found to occupy different habitats.
Inward of $\sim$100 days, planets smaller than 4$R_\oplus$ dominate, averaging to about 0.6 planet per star around FGK stars, compared to $\sim$0.04 per star for larger planets \citep[e.g.,][]{Fressin13,Christiansen15,Zhu18}.
The occurrence rate of {\it Kepler} super-Earths/mini-Neptunes (here defined as planets with radii 1--4$R_\oplus$) rises toward orbital periods of $\sim$10 days and plateaus beyond \citep[e.g.,][]{Mulders15}. Larger planets, on the other hand, rise in number toward at least $\sim$100 days \citep[e.g.,][]{Cumming08,Dong13,Santerne16}. The fact that the gas giant occurrence rate is smaller within $\sim$100 days and the break/peak appears at longer orbital periods suggests that gas giants are more likely to form farther out while Earth-sized planets are more likely to form closer in.\footnote{We note that although the number of gas giants rises toward larger distances, super-Earths/mini-Neptunes may still be the dominant population there. The global abundance of small planets is hinted at by microlensing surveys \citep[e.g.,][]{Clanton14}.}

In the theory of core accretion, whether a planet becomes a gas giant or not depends sensitively on the mass of the core \citep[e.g.,][]{Pollack96,Ikoma00,Rafikov06,Piso14,Lee14}.
Cores accrete their gaseous envelopes at rates regulated by internal cooling. Once the envelope has as much mass as the core, the gas accretion rate ``runs away'' in response to the atmosphere's self-gravity. Only those cores that are massive enough to trigger this runaway gas accretion within the disk lifetime can nucleate gas giants.
Planetesimal accretion and oligarchic growth models do make Jupiter-nucleating cores at larger distances, because more material is available for core formation beyond the ice line, but the longer dynamical timescales there
lengthen prohibitively the core coagulation timescale (e.g., the time to amass a core within a local feeding zone is on the order of Gyrs beyond $\sim$5 au; see \citealt{Goldreich04}, their equation 56).

This timescale problem is addressed by the theory of pebble accretion where particles marginally coupled to the gas (i.e., Stokes number of order unity) can be accreted rapidly to form multi-Earth-mass cores in timescales as short as $10^4$ years \citep[e.g.,][]{Ormel10,Lambrechts12}. In this scenario, planets grow to the ``pebble isolation mass'', where they start to strongly perturb the surrounding gas to create pressure maxima that barricade the cores from accreting more pebbles \citep{paardekooper04, Paardekooper06dust}. This mass rises with distance as it scales with the disk aspect ratio $h/r$ \citep[see][their equation 12]{Lambrechts14}, recovering the feature that more massive cores form at larger distances.

Pebble accretion may naturally explain the dichotomy between the inner super-Earths and outer gas giants if planets do not migrate, but migration due to the gravitational interaction with the circumstellar disks cannot be ignored. Under Type I migration \citep[e.g.,][]{Kley12}, the migration timescale for Jupiter-nucleating cores at 1 au is only about $10^5$ years. Unimpeded, migration tends to produce tightly packed planetary systems containing both large and small planets near the inner edges of disks \citep{Ogihara15}. The absence of such a pile-up in observations suggests most planetary systems undergo little to no migration \citep{Lee17}. The challenge is therefore in stopping Type I migration.

Past studies \citep[e.g.,][]{Ali-dib17} considered a switch from Type I to Type II migration as a means to slow down the wholesale migration, but gap-opening may be considerably more difficult for migrating planets  \citep[e.g,][]{Malik15}. On top of that, Type II migration rate is still under debate, as some recent simulations find that it is independent of the disk viscous flow rate \citep{Duffell14,Durmann15}. Another way to stop fast migration is to invoke planet traps \citep[e.g.,][]{Hasegawa11,Bitsch15-planet,Coleman16a,Coleman16b,brasser17} --- points in disks where the local temperature and density gradients generate corotation torques that exactly balance the Lindblad torques on the planets. To sustain the corotation torque, viscous diffusion in the disk needs to be on a level of $\alpha \gtrsim 10^{-3}$ for super-Earths near 1 au \citep{Masset10,Paardekooper11}, where $\alpha$ is the Shakura-Sunyaev parameter \citep{alpha_vis}. Gas giants can also be stopped at $\gtrsim$1 au in photoevaporative disks if the inner disk decouples from the outer disk before the majority of the planetary cores migrate inside of the wind-launching radius \citep{Alexander12,Coleman16b}.
While planet traps can aid the formation of cold Jupiters, these models find it a challenge to simultaneously reproduce sub-au super-Earths.

In this paper, we present an alternative hypothesis whereby the disk feedback torque in inviscid (or nearly laminar) disks halt the migration of planetary cores close to their initial locations.
We begin our discussion with an overview of disk-planet interactions in inviscid disks and simple calculations to predict where gas giants are more likely to appear.

\subsection{Gas Giant Formation in Inviscid Disks}

As a migrating planet repels disk material away from its orbit, gas piles up ahead and depletes behind.
In disks with sufficiently low viscosity ($\alpha \lesssim 10^{-4}$; \citealt{Li09}),
these structures are not smoothed away and can
exert a feedback torque on the planet that slows migration or even brings it to a halt \citep[][]{Hourigan84,Ward89,Rafikov02}. \citet{Rafikov02} shows that feedback can stop migration when the planet's mass reaches:
\begin{align}
\nonumber
M_{\rm fb} &= 4 M_\oplus \left(\frac{h_{\rm p}/\rp}{0.035}\right)^3 \left(\frac{\Sigma_{\rm p} \rp^2/M_*}{10^{-3}}\right)^{5/13} \\
\label{eq:M_fb}
&\sim 0.3 M_{\rm thermal} \left(\frac{\Sigma_{\rm p} \rp^2/M_*}{10^{-3}}\right)^{5/13} \, ,
\end{align}
where $\rp$ is the radial position of the planet, $h_{\rm p}$ is the disk scale height, $\Sigma_{\rm p}$ is the disk surface density, $M_*$ is the host star's mass, $M_{\rm thermal} = (h/r)^3 M_*$ is the disk ``thermal mass'', and the subscript ``p'' denotes values evaluated at the planet's position. Although migration may not stop immediately after reaching $M_{\rm fb}$ --- calculations by \citet{Rafikov02} assume steady state, which takes time to establish --- numerical simulations have verified that the migration of a super-Earth, or even a system of super-Earths, is orders of magnitude slower when disk feedback is accounted for \citep[][]{Li09,Yu10,Fung17b}.

For cores of $M_{\rm fb}$ to nucleate gas giants, they need to be massive enough to trigger runaway gas accretion within the lifetime of the natal disk. Cores can accrete as much gas as they can cool. The rate at which the envelope can cool is governed by the conditions at the innermost radiative-convective boundary (rcb); in particular, the temperature $T_{\rm rcb}$ and the density $\rho_{\rm rcb}$ at the rcb --- and therefore the opacity at the rcb.
The density at the rcb $\rho_{\rm rcb}$ is controlled by the adiabat of the inner convective zone. In this inner region, energy is spent dissociating ${\rm H}_2$ molecules. The adiabatic index $\gamma_{\rm ad}$ is driven close to 1, falling below $4/3$ and creating a centrally concentrated mass profile. Consequently, $\rho_{\rm rcb}$ is determined by the envelope mass, the core mass and radius, the adiabatic index, and $T_{\rm rcb}$ \citep[see][their equation 11]{Lee15}. What $T_{\rm rcb}$ is depends on whether the envelope is dusty (i.e., dust grains are small enough that they contribute to the total opacity) or dust-free (i.e., dust grains do not contribute to the total opacity and all metallic species are in gaseous form).

For dusty envelopes, the rcb is set by the ${\rm H}_2$ dissociation front so that $T_{\rm rcb} = 2500$ K irrespective of outer nebular conditions.
The runaway mass $M_{\rm run}$ is therefore constant with orbital distance \citep[][]{Lee14,Lee15}:
\begin{equation}
\label{eq:M_crit_dust}
    M_{\rm run,dust} = 10\,M_\oplus\left(\frac{t_{\rm disk}}{3\,{\rm Myr}}\right)^{-0.2}\left(\frac{T_{\rm rcb}}{2500\,{\rm K}}\right)^{2.8}.
\end{equation}
In deriving the above equation, we have used interstellar medium-like grain size distribution with solar metallicity \citep[see][for more detail]{Lee14}.
Although we do not show here explicitly,
the runaway mass features a non-monotonic behavior with respect to metallicity $Z$: $M_{\rm run}$ rises with $Z$ until $Z \sim 0.2$, beyond which larger mean molecular weight effects faster envelope contraction so that $M_{\rm run}$ drops with even higher $Z$ \citep[see][their Figure 1]{Lee16}.

For dust-free envelopes, the outer radiative layer becomes isothermal so that $T_{\rm rcb}$ is set by the outer temperature $T_{\rm out}$ (it can be either the nebular temperature or the stellar irradiation temperature). Farther from the star, the rcb cools down and the rcb opacity drops as the ro-vibrational modes of gaseous molecules freeze out. The envelope becomes more transparent and cools more rapidly; the runaway mass decreases with orbital distance (\citealt{Lee15}; see also \citealt{Piso15}, \citealt{Inamdar15}):
\begin{equation}
\label{eq:M_crit_df}
M_{\rm run,df} =
1.6\,M_\oplus\left(\frac{t_{\rm disk}}{3\,{\rm Myr}}\right)^{-0.4}\left(\frac{T_{\rm rcb}}{400\,{\rm K}}\right)^{1.5}.
\end{equation}
Although $T_{\rm rcb} = T_{\rm out}$, we do not explicitly state so to highlight the importance of the rcb conditions in the cooling of the envelope. Note how neither $\rho_{\rm rcb}$ nor $T_{\rm rcb}$ (and by extension the opacity at the rcb) depends on the outer nebular density.

\begin{figure}[]
\includegraphics[width=0.99\columnwidth]{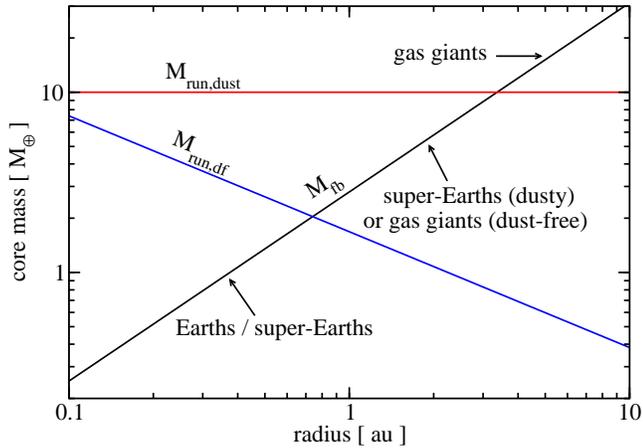}
\caption{The migration stopping mass $M_{\rm fb}$ (\eqnref{eq:M_fb}), runaway mass for dusty atmospheres $M_{\rm run,dust}$ (\eqnref{eq:M_crit_dust}), and runaway mass for dust-free atmospheres $M_{\rm run,df}$ (\eqnref{eq:M_crit_df}) as functions of distance to the star. For the disk profile, we assume a minimum mass solar nebula (described in \secref{sec:num_hydro}). More massive planetary cores stop at larger radii, and so the cores that can undergo runaway accretion are naturally separated from those that cannot. The division between these two types of planets lies around 1 au.}
\label{fig:masses}
\end{figure}

Where $M_{\rm fb}$ and $M_{\rm run}$ intersect marks a point of separation between the inner super-Earths/mini-Neptunes and the outer gas giants, as shown in \figref{fig:masses}.
We will test the validity of this idealized picture by constructing an analytic model of planet formation inviscid disks in \secref{sec:evo} and test their accuracy with numerical simulations in \secref{sec:numerics}. We present our results in \secref{sec:results} and discuss the implications in \secref{sec:conclude}.

\section{Three Stages of Formation: Deriving the Mass vs Position Relation}
\label{sec:evo}

We describe planet formation in three stages: core formation, migration, and gas accretion.
Under pebble accretion, planetary cores grow in mass until they become massive enough to create pressure maxima that can trap incoming pebbles (stage 1).
These ``pebble isolation mass'' cores undergo type I migration until the disk feedback torque halts them (stage 2).
After they come to a full stop, they spend the remainder of the disk lifetime, which lasts millions of years, accreting gas (stage 3).
Below we describe in more detail how we model each stage and discuss the circumstances in which the distinction between stages blurs.

\subsection{Stage I: Core growth and Type I Migration}
\label{sec:s1}

Cores grow in mass by accreting nearby solids. The rate of solid accretion is a strong function of the stopping time (i.e., how well-coupled the solids are to the gas, parametrized as the Stokes number St). Particles that are marginally coupled to the gas (St $\sim$ 1) are the easiest to capture: aerodynamic drag can damp away the initial kinetic energy of the incoming particles and effectively increase the accretion cross section (this is equivalent to the so-called rapid pebble accretion in the literature; see, e.g., \citealt{Lambrechts12}, see also \citealt{Ormel10} for more general discussions). Particles that are too well-coupled (St $\ll 1$) to the gas can only be accreted as much as gas would be accreted. Particles that are too decoupled (St $\gg 1$) from the gas can only be accreted as much as the core's gravity allows (this is equivalent to the traditional, gas-free planetesimal accretion).

Depending on the grain size distribution (which is poorly-constrained) and the disk temperature profile, the rate of accretion can vary by orders of magnitude. To simplify our model, we parametrize the solid accretion rate as:
\begin{equation}
\label{eq:M_dot}
\dot{\Mp} = \Mp / t_{\rm peb} \, ;
\end{equation}
equivalently, $\Mp(t)=M_0~{\rm exp}(t/t_{\rm peb})$, where $t_{\rm peb}$ is a constant.
While our prescription of core mass growth does not strictly distinguish between pebble accretion and the traditional planetesimal accretion, we envision the kind of particles that are being accreted are marginally coupled to the gas. Such particles are expected to be trapped at pressure maxima. Later in this section, we will describe how the core is expected to stop growing once it is sufficiently massive to create a pressure bump just outside of its orbit.

While the planetary core grows in mass, through its gravitational interaction with the disk, it also undergoes migration. In the linear regime where the planet's mass is too low to affect its surrounding disk structure, this is described by Type I migration. The Type I drift rate is:
\begin{equation}
\label{eq:typeI_1}
\frac{\dot{\rp}}{\rp} = -2C \Omega_{\rm p}\frac{\Sigma \rp^2}{M_*}\frac{\Mp}{M_*}\left(\frac{h_{\rm p}}{\rp}\right)^{-2}\, ,
\end{equation}
where $C$ is a constant of order unity. In general, $C$ can take on a range of values \citep[e.g.][]{Casoli09,Paardekooper10,Paardekooper11,Jimenez17}, but is generally positive in isothermal disks. Here we choose $C=2$ given by the three dimensional simulations of \citet{Fung17a}. The radial dependence in this rate is related to the disk's local density and temperature. For disks following power-law profiles
\begin{align}
\label{eq:sigma}
\Sigma &= \Sigma_0 \left(\frac{r}{r_0}\right)^{-a} \, ,\\
\label{eq:h_r}
\frac{h}{r} &= \frac{h_0}{r_0} \left(\frac{r}{r_0}\right)^{b} \, ,
\end{align}
where the subscript $0$ indicates quantities evaluated at $t = 0$, \eqnref{eq:typeI_1} can be rewritten as:
\begin{equation}
\label{eq:typeI_2}
\frac{\dot{\rp}}{r_0} = -4 \left(\frac{\Sigma_0 r_0^2}{M_*}\right) \left(\frac{h_0}{r_0}\right)^{-2} \left(\frac{\rp}{r_0}\right)^{1/2-a-2b} \left(\frac{M_0}{M_\star}\right)e^{t/t_{\rm peb}} \Omega_0 \, .
\end{equation}
The planet's ``formation path'' during this first stage (its trajectory on a mass vs position plot) is obtained by integrating the above equation:
\begin{equation}
\label{eq:stageI}
\Mp(\rp) = M_0 + \frac{M_*}{4c} \left(\frac{\Omega_0^{-1}}{t_{\rm peb}}\right) \left(\frac{\Sigma_0 r_0^2}{M_*}\right)^{-1} \left(\frac{h_0}{r_0}\right)^{2} \left(1-\left[\frac{\rp}{r_0}\right]^c\right) \, ,
\end{equation}
where $c=a+2b-1/2$.

The influx of pebbles, and therefore the planet's core growth, can be stopped if a local pressure maximum is present in the disk. For a sufficiently massive planet, planetary torques can repel disk material from the planet's orbit, and the evacuated material will form two walls on both sides of planet that can grow to become pressure maxima. \citet{Lambrechts14} found that this pebble isolation mass $M_{\rm iso}$ is $\sim0.5 M_{\rm thermal}$.

Migration prevents gap opening if the planet drifts too quickly compared to the time it needs to open a gap; this has been demonstrated numerically by \citet{Malik15}. This problem is particularly severe for sub-thermal ($\Mp<M_{\rm thermal}$) planets which require thousands of orbits to excite order unity changes in the disk profile, as we will see in \secref{sec:results}. On the other hand, if migration is stopped, or at the very least slowed down sufficiently, even very low mass planets can open disk gaps, provided disk viscosity is low \citep[][]{Duffell13,Fung14}. Therefore, the key question is not how massive the planet is, but rather when does the migration stop.

Disk feedback provides a natural mechanism for stopping planet migration in low viscosity disks. Once the planet grows to $M_{\rm fb}$ (c.f.~Equation \ref{eq:M_fb}), migration starts to slow down due to gas piling up ahead of the planet, and then a pebble trap outside of the planet's orbit can begin to form. We would therefore expect the pebble isolation mas $M_{\rm iso}$ to be set by the feedback mass $M_{\rm fb}$.

\begin{figure}
\includegraphics[width=0.99\columnwidth]{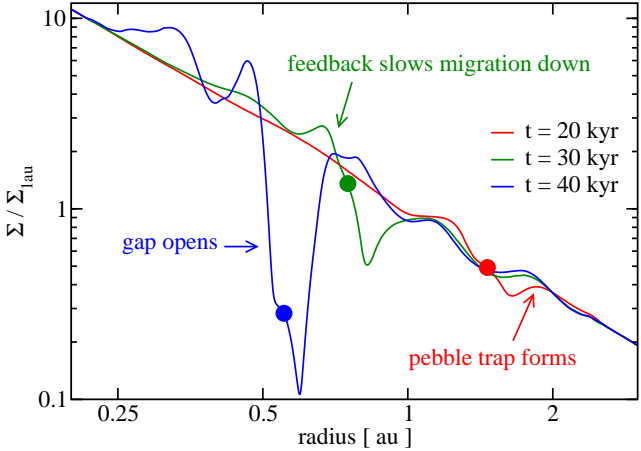}
\caption{Surface density profiles from model \#3 at 3 different epochs. The positions of the planet are marked with solid circles. At 20 kyr, the planet has grown to 6.6 $M_{\oplus}$ and perturbs the disk strongly enough to create a local pressure maximum that traps pebbles, marking the end of core growth. The planet continues to migrate inward while a feedback pile-up of gas builds ahead of it. For the next 10 kyr, the planet is gradually brought to a halt by disk feedback and begins to open a gap.}
\label{fig:den_evo}
\end{figure}

To fully stop a planet from drifting in, perturbations in the disk need to be order-unity to balance the inner and outer Lindblad torques. It is much easier to trap pebbles, which are expected to halt at any pressure maximum. See \figref{fig:den_evo} for example. Cores likely do not grow past $M_{\rm iso} \sim M_{\rm fb}$ but they will continue to migrate for some time until the pile-up of disk material ahead of the planet grows strong enough.
The next stage of planet formation features this transient migration.

\subsection{Stage II: Migration Feedback}
\label{sec:s2}

In the second stage, the planet has grown to $M_{\rm iso}\sim M_{\rm fb}$. It no longer grows in mass but continues to migrate in until the torque from the pile-up of disk material becomes strong enough to halt migration.
Our goal is to find out where the cores stop given their masses, initial locations, and the disk profiles.
The distance the core travels during this stage can be written as:
\begin{equation}
\frac{\Delta r}{\rp} = \frac{\dot{\rp}}{\rp}~t_{\rm delay} \, .
\label{eq:dr_delay}
\end{equation}
where $t_{\rm delay}$ is the time it takes for the planet-stopping perturbations to grow. Determining $t_{\rm delay}$ is nontrivial; the planet-disk interaction during this stage is a dynamic, non-linear process.
Nonetheless, we make some analytic estimates here and compare them to numerical calculations in Section \ref{sec:results}.

Despite the glaring difference between disk feedback and gap opening --- the former produces a pile-up while the latter makes a clearing --- they likely operate on a similar timescale: the time it takes a planet to build order-unity perturbations in the disk through the torque it exerts. We take $t_{\rm delay} \sim t_{\rm gap}$ where $t_{\rm gap}$ is the gap opening timescale \citep{Rafikov02}:
\begin{align}
\nonumber
t_{\rm gap} \sim \frac{t_{\rm cross}}{\lambda_{\rm t}} &\propto \left(\frac{\Mp}{M_\star}\right)^{-14/5} \left(\frac{h_{\rm p}}{\rp}\right)^{37/5} \Omega_{\rm p}^{-1} \\
\label{eq:t_gap_rafikov}
&\propto \left(\frac{\Mp}{M_{\rm thermal}}\right)^{-14/5} \left(\frac{h_{\rm p}}{\rp}\right)^{-1} \Omega_{\rm p}^{-1} \, ,
\end{align}
where $t_{\rm cross}$ ($t_0$ in the notation of \citet{Rafikov02}; see their equation 37) is the timescale for gas to drift across the planet's gap in the rest frame of the planet, and $\lambda_{\rm t}$ (same notation as \citet{Rafikov02}; see their equation 39) is a parameter that quantifies the amount of angular momentum deposition due to weak shocks.

Setting $t_{\rm delay} \sim t_{\rm gap}$ and substituting Equations \ref{eq:typeI_1} and \ref{eq:t_gap_rafikov} into \eqnref{eq:dr_delay}, we get:
\begin{align}
\nonumber
\frac{\Delta r}{\rp} &\propto \frac{\Sigma \rp^2}{M_*} \left(\frac{\Mp}{M_*}\right)^{-9/5} \left(\frac{h_{\rm p}}{\rp}\right)^{27/5} \\
&\propto \frac{\Sigma \rp^2}{M_*} \left(\frac{\Mp}{M_{\rm thermal}}\right)^{-9/5} \, .
\end{align}
During $t_{\rm delay} \sim t_{\rm gap}$, planets do not grow in mass so we set $M_{\rm p} = M_{\rm fb}$, and the initial location
$\rp$ is set by the location of the cores $r_{\rm fb}$ when they grow to $M_{\rm fb}$.
We arrive at the following scaling relation:
\begin{equation}
\label{eq:drift_dis}
\frac{\Delta r}{r_{\rm fb}} \propto \left(\frac{\Sigma r_{\rm fb}^2}{M_*}\right)^{4/13}  \, .
\end{equation}
The final position of the planet after feedback is able to fully halt migration is:
\begin{equation}
\label{eq:rfinal}
r_{\rm final} = r_{\rm fb} + \Delta r = r_{\rm fb}\left(1-0.5\left(\frac{\Sigma r_{\rm fb}^2 / M_*}{10^{-4}}\right)^{4/13}\right) \,
\end{equation}
where the coefficient 0.5 was computed numerically (see \secref{sec:results}). Because we have used a locally estimated $t_{\rm delay}$ and migration rate, this formula for $r_{\rm final}$ works best when it is within order-unity of $r_{\rm fb}$.

\subsection{Stage III: Gas Accretion}
\label{sec:s3}

The rate at which cores accrete gas is mediated by the rate at which the gas can cool.
During the initial stage of core growth, heating by solid accretion overwhelms the atmosphere's ability to cool and so the cores barely build their gaseous envelopes. The rate of heating generated by the release of gravitational energy as solids fall onto the surface of the core is
\begin{equation}
    \frac{GM_{\rm core}^2}{R_{\rm core}t_{\rm peb}} \sim 2\times 10^{29}\,{\rm ergs\,s^{-1}}\left(\frac{M_{\rm core}}{5\,M_\oplus}\right)^{5/3}\left(\frac{t_{\rm peb}}{10^4\,{\rm yrs}}\right)^{-1}
    \label{eq:Lheat}
\end{equation}
where $R_{\rm core} = 1.6R_\oplus(M_{\rm core}/5\,M_\oplus)^{1/3}$ is the radius of the core. For typical super-Earth masses and pebble accretion rates,
this heating rate is orders of magnitude larger than the cooling rate of the envelope whether the dust grains dominate the opacity,
\begin{equation}
    L_{\rm cool,dusty} \sim 2\times 10^{26}\,{\rm erg\,s^{-1}}\left(\frac{M_{\rm gas}/M_{\rm core}}{0.01}\right)^{-1.5}\left(\frac{M_{\rm core}}{5\,M_\oplus}\right)^6
    \label{eq:Lcool_dusty}
\end{equation}
or not,
\begin{equation}
    L_{\rm cool,df} \sim 2\times 10^{28}\,{\rm erg\,s^{-1}}\left(\frac{T_{\rm rcb}}{370\,{\rm K}}\right)^{-4.5}\left(\frac{M_{\rm gas}/M_{\rm core}}{0.01}\right)^{-1.6}\left(\frac{M_{\rm core}}{5\,M_\oplus}\right)^{4.6}
    \label{eq:Lcool_df}
\end{equation}
where we take equation 13 from \citet{Lee15} for the cooling luminosity in combination with their equation 18 for dusty and equation 23 for dust-free accretion. The envelope mass is expressed as $M_{\rm gas}$. Even at thermal equilibrium (i.e., heating is balanced by cooling), the expected mass fractions are only $M_{\rm gas}/M_{\rm core} \sim 10^{-4}$ for dusty and $M_{\rm gas}/M_{\rm core} \sim 10^{-3}$ for dust-free envelopes.

Gas accretion therefore begins once solid accretion ends. In the absence of heating, the core is free to accrete and build its gaseous envelope. We adopt the semi-analytic scaling relationship from \citet[][see also \citealt{Ginzburg16}]{Lee15} who provide the time evolution of envelope mass fraction for both dusty and dust-free accretion:
\begin{equation}
\label{eq:mgas_dusty}
M_{\rm gas,dusty} \simeq 0.5\,M_\oplus \left(\frac{t}{3 \,{\rm Myr}}\right)^{0.4} \left(\frac{M_{\rm core}}{5M_\oplus}\right)^{2.7} \left(\frac{f_\Sigma}{0.1}\right)^{0.12}
\end{equation}
and
\begin{equation}
\label{eq:mgas_df}
   M_{\rm gas,df} \simeq 1.3\,M_\oplus \left(\frac{t}{3\,{\rm Myr}}\right)^{0.4}\left(\frac{T_{\rm rcb}}{800\,{\rm K}}\right)^{-1.5} \left(\frac{M_{\rm core}}{5M_\oplus}\right)^2 \left(\frac{f_\Sigma}{0.1}\right)^{0.12}
\end{equation}
where $f_\Sigma \equiv \Sigma_{\rm p}/\Sigma_{\rm mmsn}$, the density depletion factor with respect to the minimum mass solar nebula, is used to account for changes in the gas density in different disk models and depletion due to gap-opening.
The scaling relationships are modified for the weak dependence on the nebular density (see Figure 4 of \citealt{Lee16} and \S\ref{sec:intro} for a discussion).

Once the envelope mass becomes comparable to the core mass, the self-gravity of the envelope becomes significant:
stronger gravity demands faster cooling, triggering the runaway gas accretion \citep{Pollack96}.
We therefore classify any planet that gains $M_{\rm gas} \sim 0.5 M_{\rm core}$ within our assumed disk lifetime $\sim$3 Myr as gas giants. The semi-analytic scaling relationships we adopt are accurate to factors of order unity.
For models that are considered to be on the verge of (but not quite at) runaway by our analytic expressions (see notes in \tabref{tab:para}), we run additional numerical evolutionary models from \citet{Lee14} to more accurately determine their fates.

We note that stages II and III are not always distinct. Dust-free gas accretion can sometimes proceed so rapidly that the planet undergoes runaway accretion before migration stops. These extreme instances have little impact on our results however, since we are most interested in the marginal cases lying between super-Earths and gas giants.

\section{Numerical Methods}
\label{sec:numerics}

\subsection{Hydrodynamical Simulations}
\label{sec:num_hydro}

Our numerical setup borrows from \citet{Fung17b}, and we recapitulate here some of the main features. We perform 2D simulations of disk-planet interactions using the graphics processing unit (GPU) accelerated hydrodynamics code \texttt{PEnGUIn} \citep{MyThesis}. It is a Lagrangian-remap shock-capturing code with a Riemann solver that follows the piecewise parabolic method \citep{PPM}.
It has been updated to include the fast orbital advection algorithm \citep{Masset00}, which allows the code to take time steps unrestricted by the background Keplerian motion.

\texttt{PEnGUIn} solves the continuity and momentum equations:
\begin{align}
\label{eq:cont_eqn}
\frac{{\rm D}\Sigma}{{\rm D}t} &= -\Sigma\left(\nabla\cdot\mathbf{v}\right) \,,\\
\label{eq:moment_eqn}
\frac{{\rm D}\mathbf{v}}{{\rm D}t} &= -\frac{1}{\Sigma}\nabla P - \nabla \Phi \,,
\end{align}
where $\Sigma$ is the gas surface density, $\mathbf{v}$ the velocity field, $P$ the vertically averaged gas pressure, and $\Phi$ the combined gravitational potential of the star and the planet.

In polar coordinates ($r$, $\phi$) centered on the star,
\begin{align}\label{eq:potential}
\Phi = -\frac{GM_\ast}{r} - \frac{G\Mp}{\sqrt{r^2 + \rp^2 - 2r\rp\cos{\phi'} + \rs^2}} + \frac{G\Mp r\cos{\phi'}}{\rp^2} \,,
\end{align}
where $G$ is the gravitational constant, $M_\ast = 1 M_\odot$ is the stellar mass, $\rs$ the smoothing length of the planet's potential, and $\phi' = \phi-\phi_{\rm p}$ the azimuthal separation from the planet. The third term on the right is the indirect potential. We set $\rs = 0.5 h$ to approximate the vertically averaged gravitational force \citep{Muller12}.

We use a locally isothermal equation of state: $P= (k_{\rm B} T/\mu m_{\rm H})\Sigma$, where $k_{\rm B}$ is the Boltzmann constant, T is the disk temperature, $\mu=2.34$ the mean molecular weight, and $m_{\rm H}$ the hydrogen mass. The disk temperature is given a radial dependence:
\begin{equation}
T = 370~{\rm K} \left(\frac{r}{1~{\rm au}}\right)^{-\frac{3}{7}} \, ,
\label{eq:Tdisk}
\end{equation}
following the passively heated disk model of \citet{Chiang97}. Note that this translates to $b=2/7$ and $h_0/r_0 = 0.038$ if $r_0 = 1~{\rm au}$ in \eqnref{eq:h_r}. For our fiducial model, the disk surface density follows the minimum mass solar nebula (MMSN; \citealt{MMSN}):
\begin{equation}
\Sigma = \Sigma_{\rm 1au} \left(\frac{r}{1~{\rm au}}\right)^{-\frac{3}{2}} \, ,
\label{eq:Sigdisk}
\end{equation}
where $\Sigma_{\rm 1au}$ is the surface density at 1 au, and is 1700 ${\rm g~cm^{-2}}$ for MMSN. We denote this MMSN surface density profile $\Sigma_{\rm mmsn}$. For 5 of our models (\#1-5), we use the MMSN value; and for 2 additional runs (models \#6 and 7) we use 4 times lower density (\tabref{tab:para}).
The velocity field $\mathbf{v}$ initially has zero radial velocity and the azimuthal rotational frequency $\Omega$ balances gravity and gas pressure:
\begin{equation}
\Omega = \sqrt{\frac{GM_\ast}{r^3} + \frac{1}{r\Sigma}\frac{{\rm d}P}{{\rm d}r}} \, .
\end{equation}

\begin{deluxetable*}{ccccccccccc}
\tablecaption{\label{tab:setup} Model Parameters and Results}
\tablehead{Model \#&$r_{\rm start}$ & $\Sigma_{\rm 1 au}$ & $r_{\rm in}$ & $r_{\rm out}$ & $t_{\rm end}$ & $r_{\rm final}$ & $M_{\rm core, final}$ & Runaway? & Runaway?\\
 & [au] & [g $\rm cm^{-2}$] & [au] & [au] & [kyr] & [au] & [$M_\oplus$] & (dusty) & (dust-free)}
\startdata
1 & 1.0 & 1700 & 0.2  & 2.0 & 25 & 0.27 & 3.1 & No  & No \\
2 & 1.5 & 1700 & 0.2  & 3.0 & 40 & 0.44 & 4.9 & No  & Yes\\
3 & 2.0 & 1700 & 0.2  & 3.0 & 40 & 0.56 & 6.6 & No  & Yes\\
4 & 2.5 & 1700 & 0.3  & 4.5 & 50 & 0.67 & 8.3 & No\tablenotemark{a}  & Yes\\
5 & 3.5 & 1700 & 0.3  & 4.5 & 50 & 1.2  & 13  & Yes & Yes\\
\hline
6 & 2.0 & 425 & 0.4  & 4.0 & 130 & 0.90 & 5.5 & No & Yes\\
7 & 3.5 & 425 & 0.5  & 7.5 & 150 & 1.6  & 12 & Yes\tablenotemark{b} & Yes
\enddata
\tablenotetext{a}{A marginal case where the planet would undergo runaway accretion if the disk lifetime were 3.5 Myr instead of our assumed 3 Myr.}
\tablenotetext{b}{Analytic scaling relationship suggests the planet to be on the verge of runaway but numerical calculations show it to have reached envelope mass fraction of $\gtrsim$50\% by $\sim$1.5 Myrs.}
\label{tab:para}
\end{deluxetable*}

\subsubsection{Grid Parameters and Boundary Conditions}
\label{sec:num_para}

Our simulation grid is in polar coordinates, and grid cells are spaced logarithmically in radius and uniformly in azimuth. The resolution is $\Delta r/r \sim \Delta \phi \sim 0.0032$, or about 12 cells per scale height at 1 au. We simulate the full $2\pi$ in azimuth, and the radial extent goes from the inner boundary at $r_{\rm in}$ to the outer boundary $r_{\rm out}$. \tabref{tab:para} lists $r_{\rm in}$ and $r_{\rm out}$ for each model. All simulations last until planet migration completely halts. The time it takes for the planets to come to a full stop $t_{\rm end}$ is also listed in \tabref{tab:para}.

Azimuthal boundaries are periodic.
We employ fixed boundary conditions at both $r_{\rm in}$ and $r_{\rm out}$ where we attach wave killing zones. The wave killing zones are one scale height in width, and their prescription is:
\begin{equation}
\dot{X} = \frac{X - X(t=0)}{20\pi\Omega^{-1}} \left(1-\frac{d}{h}\right)^2 \, ,
\end{equation}
where $X$ represents fluid properties $P$, $\Sigma$ and $\mathbf{v}$, $d$ is the distance to the boundary, and $\Omega$ and $h$ are the orbital frequency and disk scale height evaluated at the boundary.

\subsubsection{Planet Evolution}
\label{sec:num_planet}

Following \citet{Fung17b}, we integrate the planet's motion using a kick-drift-kick leapfrog scheme, with the drift step occurring synchronously with the hydrodynamics step, and we treat the planet's position as linear in time within a hydrodynamics step. The force on the planet exerted by the disk is computed by summing over the gravitational force from all the mass elements in the simulation grid, with the ``background'' axisymmetric component of the disk surface density subtracted off. Since the disk does not feel its own gravity in our model, eliminating the axisymmetric disk-planet forcing ensures a more consistent motion between the planet and its neighboring disk elements. To ensure numerical stability, we have verified that the total force from within the planet's smoothing length $\rs$, which is generally much larger than its Hill radius, is negligible compared to the rest of the disk.

Cores are initially placed at $r_{\rm start} \in [1,3.5]$ au (see \tabref{tab:para}) with masses set to $\Mp = 1 M_\oplus$. Their mass growth follows \eqnref{eq:M_dot}. Whenever a local pressure maximum is detected outward to the planet's orbit, we set the growth rate to zero.
The mass doubling timescale $t_{\rm peb}$ is set to $10^4$ yrs; we choose a short $t_{\rm peb}$ so that our simulations can last for at least a few $t_{\rm peb}$.
We discuss in \secref{sec:results} the impact of adopting more realistic scheme of solid accretion.

Once the cores halt completely, we stop the hydrodynamical simulations. We estimate the amount of gas each core will accrete within the disk lifetime (assumed to be 3 Myrs) using Equations \ref{eq:mgas_dusty} and \ref{eq:mgas_df}.
The amount of gas a core accretes depends weakly on the nebular density, parametrized as the depletion factor $f_\Sigma$ with respect to the background unperturbed gas disk. At the end of the hydrodynamical simulations, $f_\Sigma \sim$0.03--0.3 (for models \#1--5; it is $\sim$4 times lower in models \#6 and 7) in the gaps that the planets carve out (e.g., see Figue \ref{fig:den_evo}); we take $f_\Sigma = 0.1$ (0.025 for models \#6 and 7) for simplicity. As the planets grow in total (core + gas) mass, the gaps will likely deepen and $f_\Sigma$ will drop, but the weak sensitivity of the gas accretion rate on the nebular density --- the final envelope mass differs by factors of $\sim$3 over 5 orders of magnitude change in the nebular density --- assures that our estimates of the envelope mass (prior to runaway) is robust to the uncertainties in gap depths.

Planets are assumed to become gas giants once their envelope masses reach more than half the mass of their cores, at which point the envelope cooling time shortens catastrophically and runaway accretion ensues.
We do not model this phase of evolution, and only discuss how it may proceed here.
In runaway accretion, the rate of gas accretion is no longer limited by the cooling rate but by the rate at which the disk transports gas to the planet's feeding zone. In inviscid disks, gas may be delivered by planetary torques of neighboring planets \citep{Goodman01,Sari04,Fung17b}, disk winds \citep[e.g.,][]{Bai13,Gressel15,Wang17}, and the Hall effect \citep[e.g.,][]{Lesur14,Bai15,Simon15}. Additionally, hydrodynamical instabilities at planetary gap edges such as the Rayleigh instability \citep{Fung16} and Rossby wave instability \citep{Li05} should refill the gaps if they become too depleted.

For the cases where the cores are just on the verge of runaway (i.e., $0.4 < M_{\rm gas}/M_{\rm core} < 0.5$), we perform one-dimensional numerical model of gas accretion outlined in \citet{Lee14}. We have verified that the amount of gas the cores accrete during $t_{\rm end}$ (the duration of a hydrodynamical simulation until the core comes to a full-stop; see \tabref{tab:para}) is negligible.

\section{Results}
\label{sec:results}

\begin{figure}
\includegraphics[width=0.99\columnwidth]{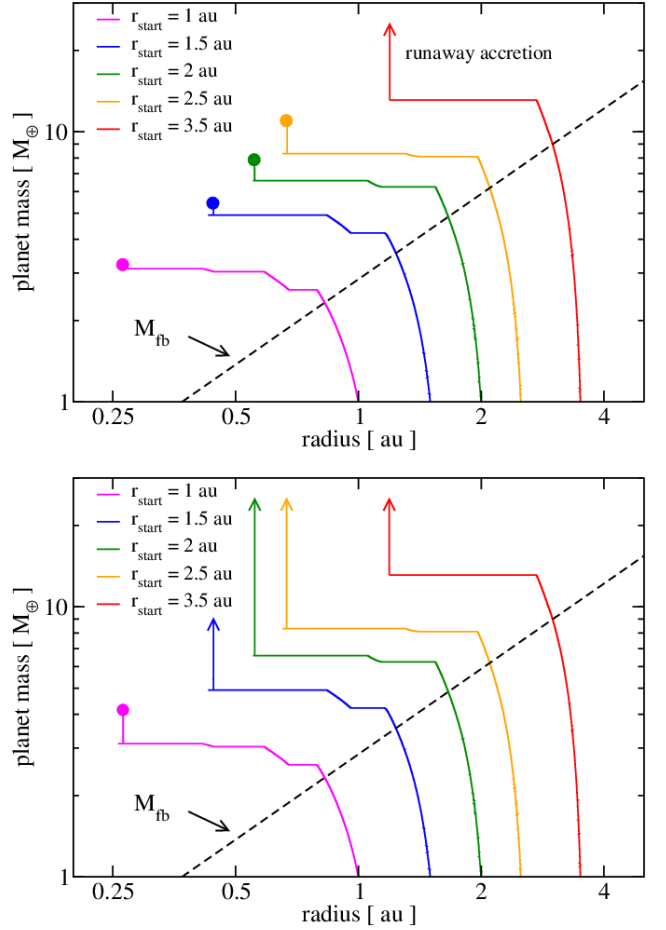}
\caption{Total planet mass (core mass + gas mass) vs radial position for models \#1-5, corresponding to $r_{\rm start} =$ 1, 1.5, 2, 2.5, and 3.5 au respectively. The top panel shows the evolution with dusty atmospheres, and bottom panel for dust-free atmospheres. The planets that reach a gas-to-core-ratio of 0.5 within 3 Myr are expected to undergo runaway gas accretion and are denoted with an upward arrow. Those that do not undergo runaway are marked with solid circles. In the dusty cases, model \#4 (orange) is a marginal case that is on the verge of runaway (see comment in \tabref{tab:para}). We find that super-Earths and gas giants are spatially separated at $\sim$0.7 au in the dusty case, and $\sim$0.3 au in the dust-free case.}
\label{fig:evo_mmsn}
\end{figure}

\begin{figure}
\includegraphics[width=0.99\columnwidth]{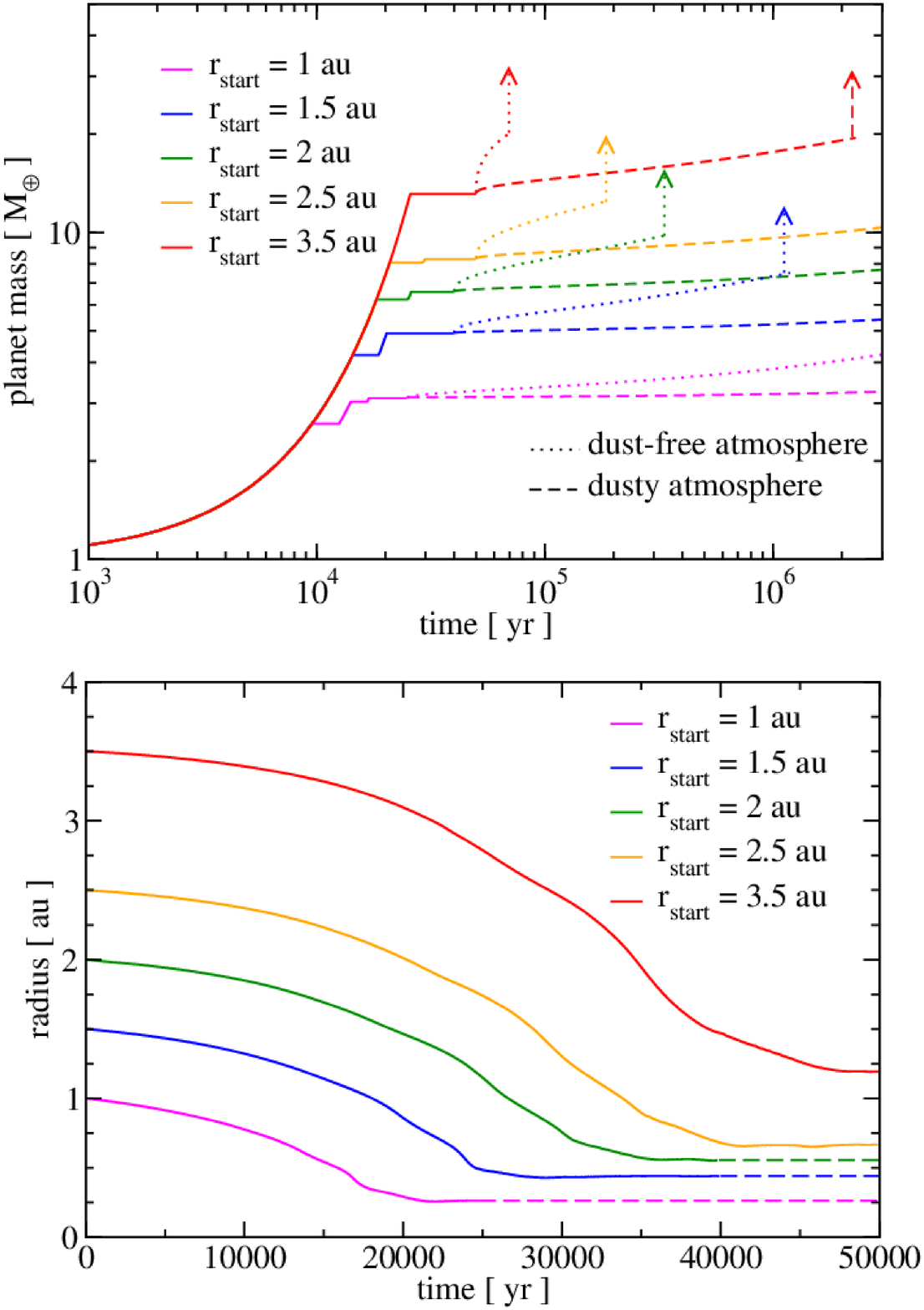}
\caption{Total planet mass (top) and radial position (bottom) as functions of time for models \#1-5. The solid curves are results from our hydrodynamical simulations, and their dashed (dotted) extensions are the analytic predictions for gas accretion in dusty (dust-free) atmospheres, given by Equations \ref{eq:mgas_dusty} and \ref{eq:mgas_df}. Vertical arrows indicate runaway gas accretion.
Cores grow over a couple of $t_{\rm peb} = 10^4$ yrs until they become massive enough to create pressure maxima outward to their orbits (\secref{sec:s1}). After that, they accrete gas for up to the disk lifetime $t_{\rm disk} =$ 3 Myr (\secref{sec:s3}). Those that reach gas-to-core mass ratio of 0.5 and above within $t_{\rm disk}$ are expected to undergo runaway gas accretion and become gas giants.
In the bottom panel, we show more clearly the behavior of planet migration in inviscid disks. Cores initially migrate inward due to Type I migration (\secref{sec:s1}) but are gradually brought to a halt by disk feedback (\secref{sec:s2}).}
\label{fig:time_mmsn}
\end{figure}

\begin{figure}
\includegraphics[width=0.99\columnwidth]{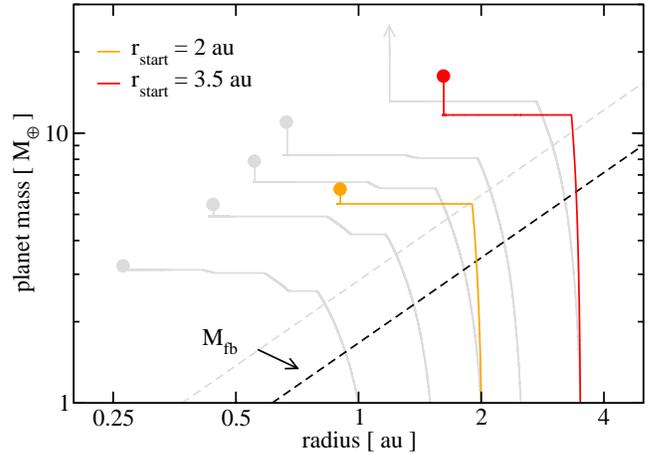}
\caption{Total planet mass (core mass + gas mass) vs radial position for models \#6-7 which have 4 times lower disk surface density than those shown in \figref{fig:evo_mmsn}. The results of models \#1--5 (top panel of \figref{fig:evo_mmsn}) are overlaid in grey for comparison. The evolution here are for dusty atmospheres, using the analytic scaling relationship given in \eqnref{eq:mgas_dusty}. Model \#7 is shown here to be on the verge of going runaway, but direct numerical calculations suggest this planet will reach a gas-to-core-ratio of 0.5 in 1.5 Myr (see notes in \tabref{tab:para}). For dust-free atmospheres, which we omit to show here, both models undergo runaway gas accretion within 3 Myrs (see \tabref{tab:para}).}
\label{fig:evo_mmsn_4}
\end{figure}

\begin{figure}
\includegraphics[width=0.99\columnwidth]{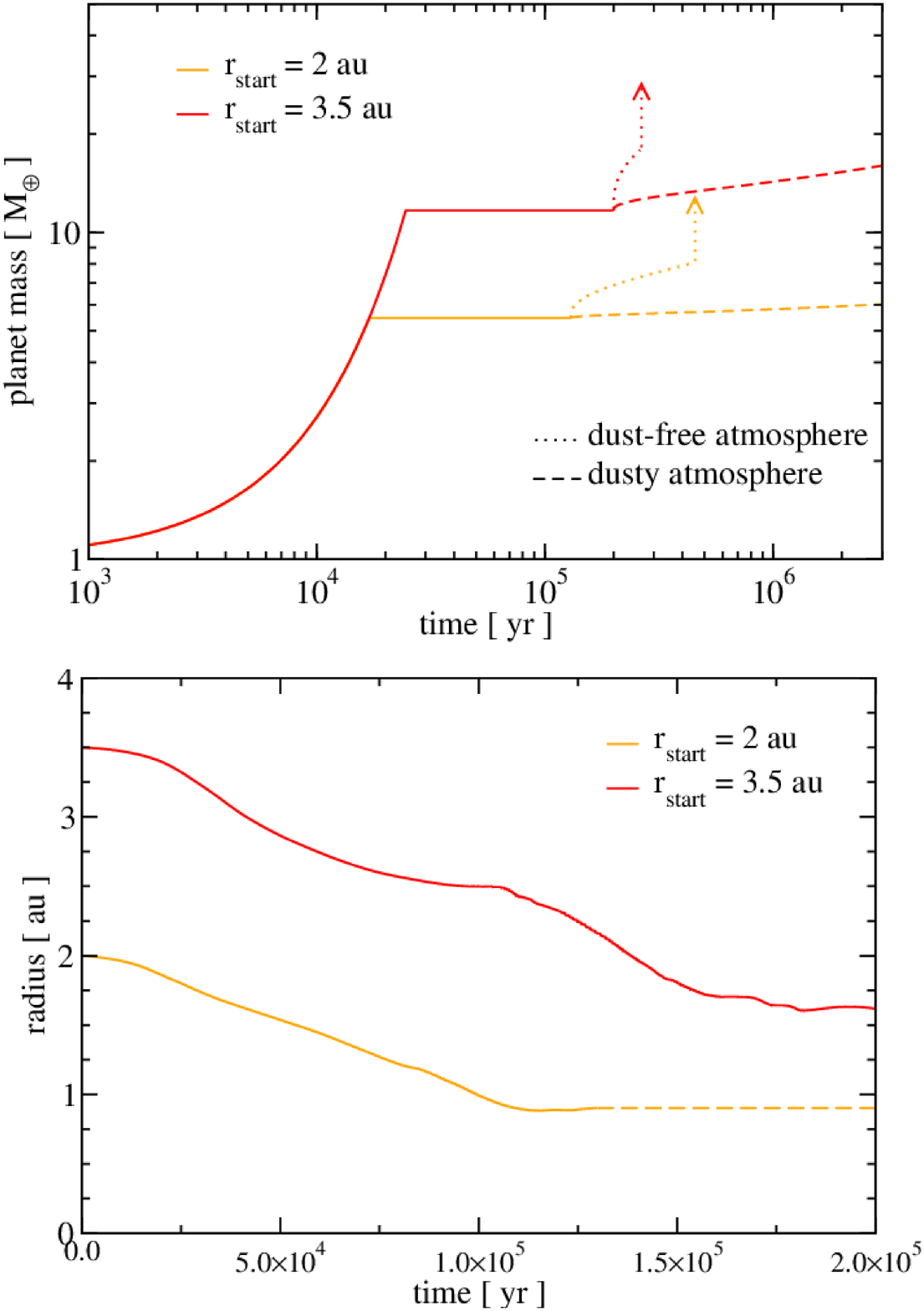}
\caption{Same as \figref{fig:time_mmsn} but for models \#6-7 which have 4 times lower disk surface density. Compared to \figref{fig:time_mmsn}, planets in less massive disks migrate more slowly and take a longer time to come to a halt. We also find that lower mass planets are stopped at larger radii, pushing the formation of gas giants to a larger radius.}
\label{fig:time_mmsn_4}
\end{figure}

Figures \ref{fig:evo_mmsn} and \ref{fig:time_mmsn} show the evolution of both the masses and orbital radii of 5 planets in an MMSN-like disk (models \#1--5).
In general, feedback masses $M_{\rm fb}$ and radii $r_{\rm fb}$ are good proxies for the pebble isolation masses and the final locations of the cores, verifying our predictions in \secref{sec:s1}. In our MMSN models, core growth stops at $\sim$1.4$M_{\rm fb}$ and migration halts at $\sim$0.3$r_{\rm fb}$; in a less massive disk (models \#6,7), we find $M_{\rm iso}\sim$1.9$M_{\rm fb}$ and $r_{\rm final} \sim$0.5$r_{\rm fb}$ (see Figures \ref{fig:evo_mmsn_4} and \ref{fig:time_mmsn_4}). With formal fits, we find
\begin{equation}
\label{eq:M_iso}
M_{\rm iso} =
\begin{cases}
4\,M_\oplus \left(\frac{r_{\rm fb}}{1~{\rm au}}\right) & \Sigma = \Sigma_{\rm mmsn} \, , \\
3\,M_\oplus \left(\frac{r_{\rm fb}}{1~{\rm au}}\right) & \Sigma = \Sigma_{\rm mmsn}/4 \, , \\
\end{cases}
\end{equation}
and
\begin{equation}
\label{eq:drift_dis_emp}
\frac{\Delta r}{r_{\rm fb}} = -0.5\left(\frac{\Sigma r_{\rm fb}^2/M_\ast}{10^{-4}}\right)^{4/13}  \, .
\end{equation}
The numerically determined $\Delta r/r_{\rm fb}$ is in broad agreement with the scaling relation given by \eqnref{eq:drift_dis}.
The corresponding $t_{\rm delay}$ is:
\begin{align}
\nonumber
t_{\rm delay} &\sim 10^4~{\rm yr} \left(\frac{\Mp}{10~M_\oplus}\right)^{-14/5} \left(\frac{h_{\rm p}/\rp}{0.035}\right)^{37/5} \left(\frac{2\pi\Omega_{\rm p}^{-1}}{1~{\rm yr}}\right)  \\
\label{eq:t_delay_emp}
&\sim 10^4~{\rm yr} \left(\frac{\Mp}{M_{\rm thermal}}\right)^{-14/5} \left(\frac{h_{\rm p}/\rp}{0.035}\right)^{-1} \left(\frac{2\pi\Omega_{\rm p}^{-1}}{1~{\rm yr}}\right)  \, .
\end{align}

In models \#1--4, we find that the cores grow episodically past $10^4$ yrs (see \figref{fig:time_mmsn}).
This corresponds to the appearance and disappearance of outer pressure bumps. Because our simulations are inviscid, pressure bumps are not erased by disk viscosity; rather, it is the planet that erases them.
Rapidly migrating planets can sometimes build a new bump just inside of the old one,
effectively ironing out these local perturbations.
Pebble accretion resumes until a new outer pressure bump appears.
This episodic accretion is not seen in lower mass disks where migration is slower (see \figref{fig:evo_mmsn_4}).

Our $M_{\rm iso}$ is about factors of 2--3 smaller compared to that found by \citet{Lambrechts14}. Their simulations differ in a few ways from ours: they simulated 3D viscous disks with $\alpha=6\times10^{-3}$ in which planets were held on fixed orbits. We simulate 2D inviscid disks with planets that migrate. A lower viscosity better preserves planet-induced disk structures, so it is expected that we find smaller $M_{\rm iso}$.

Our numerical results are largely in agreement with our analytic descriptions in \secref{sec:evo}. In particular, we correctly predict the shorter radial drift when disk mass is reduced (models \#6 and \#7). We do, however, find $M_{\rm iso}$ to be slightly larger than $M_{\rm fb}$. We postulate that the answer lies in the competition between core growth and the creation of pressure maxima.
If the core grows too fast before it has a chance to create pressure bumps, its final mass will overshoot $M_{\rm fb}$ significantly. We may find $M_{\rm iso}$ to approach $M_{\rm fb}$ as we lengthen $t_{\rm peb}$. To test this numerically requires longer simulations that are beyond the capability of our current computational resources, but may be possible in the future.

\subsection{Final Locations of Gas Giants vs.~Super-Earths}
\label{sec:final_loc}

If the cores accrete dusty gas, gas giants are found to form outside of $\sim$0.7 au; super-Earths form inside. This dividing line shrinks to $\sim$0.3 au for dust-free gas accretion. When we lower the overall disk mass, the gas giant / super-Earth division moves to a larger radius. This is mainly due to two effects: in less massive disks, $M_{\rm iso}$ is smaller and $\Delta r$ is shorter; in other words, cores form smaller and halt closer to where they start.
Gas accretion rates also drop slightly in less massive disks so a larger core mass is required to nucleate gas giants, but this is a weak effect, as shown in Equations \ref{eq:mgas_dusty} and \ref{eq:mgas_df}.

\begin{figure}
\includegraphics[width=0.99\columnwidth]{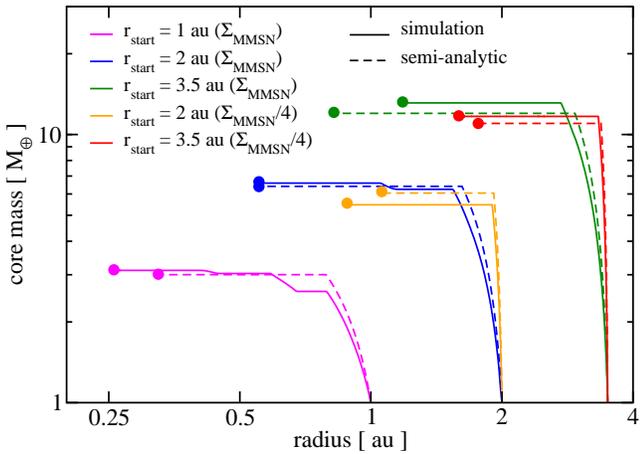}
\caption{Core mass vs.~radial position from simulations (solid lines) and semi-analytic calculations (dashed lines; Equations \ref{eq:stageI}, \ref{eq:rfinal}, and \ref{eq:M_iso}). Solid circles mark the final core masses and positions of the planets after both the pebble accretion and migration have stopped. The semi-analytic calculations recover the core masses to within 10\%, and their positions to within 30\%.}
\label{fig:ana}
\end{figure}

Combining our numerical results and analytic model (\secref{sec:evo}), we can write down a general expression for the division between inner super-Earths and outer gas giants.
We first determine the final locations of the cores as a function of their masses by substituting \eqnref{eq:M_iso} into \eqnref{eq:rfinal}:
\begin{align}
    r_{\rm final} &\sim 1\,{\rm au}\left(\frac{M_{\rm iso}}{4\,M_\oplus}\right)\left(\frac{\Sigma_{1{\rm au}}}{1700\,{\rm g\,cm^{-2}}}\right)^{-5/13} \nonumber \\
    &\times \left[1-0.6\left(\frac{M_{\rm iso}}{4\,M_\oplus}\right)^{2/13}\left(\frac{\Sigma_{1{\rm au}}}{1700\,{\rm g\,cm^{-2}}}\right)^{42/169}\right] \, ,
    \label{eq:rfinal_num}
\end{align}
assuming $M_\star=M_\odot$ and disk profiles as shown in Equations \ref{eq:Tdisk} and \ref{eq:Sigdisk}.
Because the two branches in \eqnref{eq:M_iso} are similar, for simplicity we assume $M_{\rm iso} \sim 1.4 M_{\rm fb} \sim 4 M_\oplus\,(r_{\rm fb}/1\,{\rm au})\,(\Sigma_{1{\rm au}}/1700\,{\rm g\,cm^{-2}})^{-5/13}$. In \figref{fig:ana}, we adopt both branches and show that the semi-analytic approximations agree with the numerically determined $M_{\rm iso}$ and $r_{\rm final}$ to within 10\% and 30\%, respectively.

\begin{figure}
\includegraphics[width=0.99\columnwidth]{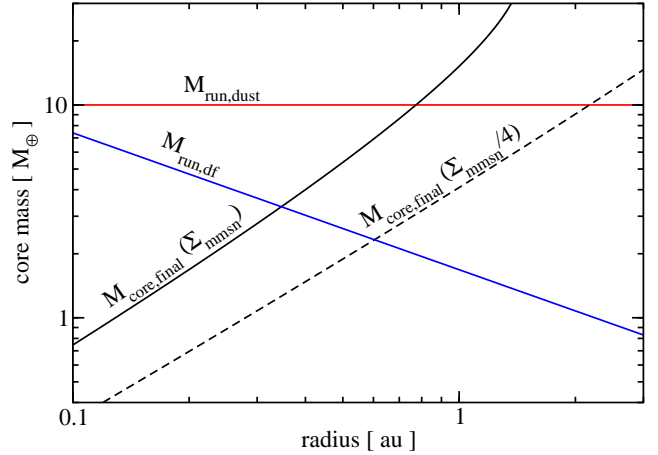}
\caption{Similar to \figref{fig:masses}, but with $M_{\rm fb}$ replaced by the final core mass $M_{\rm core,final} = M_{\rm iso}$ as a function of the final location $r_{\rm final}$ (\eqnref{eq:rfinal_num}). When $M_{\rm iso}$ is higher than $M_{\rm run}$ (assuming either a dusty (``dust'') or dust-free (``df'') atmosphere), gas giants are expected to form. In general, gas giants emerge preferentially at large distances.}
\label{fig:masses_final}
\end{figure}

To calculate the division radius $r_{\rm div}$ between the inner super-Earths and the outer gas giants, we let $M_{\rm iso}$ in \eqnref{eq:rfinal_num} equal the runaway masses $M_{\rm run,dust}$  (\eqnref{eq:M_crit_dust}) and $M_{\rm run,df}$ (\eqnref{eq:M_crit_df}):
\begin{equation}
    r_{\rm div} \sim
    \begin{cases}
    0.8\,{\rm au}\left(\frac{t_{\rm disk}}{3\,{\rm Myr}}\right)^{-0.2}\left(\frac{1700\,{\rm g\,cm^{-2}}}{\Sigma_{1{\rm au}}}\right)^{0.4} & {\rm dusty\,accretion} \, , \\
    0.3\,{\rm au}\left(\frac{t_{\rm disk}}{3\,{\rm Myr}}\right)^{-0.2}\left(\frac{1700\,{\rm g\,cm^{-2}}}{\Sigma_{1{\rm au}}}\right)^{0.2} & {\rm dust-free\,accretion} \, .
    \end{cases}
    \label{eq:rdiv}
\end{equation}
Gas giants appear at larger distances in less massive disks because migration is slower in these disks.
Figure \ref{fig:masses_final} summarizes our results.
Our calculation of $r_{\rm div}$ is weakly dependent on the disk lifetime: $r_{\rm div}$ changes by only 30\%--60\% for an order of magnitude uncertainty in $t_{\rm disk}$.

\section{Summary and Discussion}
\label{sec:conclude}

Both radial velocity and transit surveys reveal that the population of gas giants around FGK stars rises toward longer orbital periods. Inside orbital periods of $\sim$100 days, gas giants appear only around $\sim$3\% of stars, compared to super-Earths' $\sim$60\%. These observations suggest that the outer regions of protoplanetary disks provide more favorable formation condition of gas giants.

Using hydrodynamic simulations and semi-analytic calculations, we have shown that a combination of pebble accretion and disk feedback in inviscid disks can naturally explain why gas giants likely form cold and stay cold while the inner planetary systems are dominated by super-Earths/mini-Neptunes. Three effects conspire to make the outer disk regions natural breeding grounds of gas giants: 1) cores can grow more massive under pebble accretion; 2) cores barely migrate under disk feedback; and 3) gas accretion can proceed more rapidly. That the cores undergo stunted migration is important. Under classical Type I migration, all the cores would have piled up at the inner edge of the disk and the mass ordering set by $M_{\rm iso}$ would be mixed up as more massive cores migrate
faster.\footnote{By contrast, {\it Kepler} planets in a given system are found to be ordered in mass \citep{Millholland17}.}
We find the division between the inner super-Earths and outer gas giants, $r_{\rm div}$, to lie between 0.3 and 0.8 au, which we consider accurate to within order-unity. There is a large parameter space that we have not yet explored. Below we discuss some of them and use them to motivate future investigations.

\subsection{Pebble Accretion Timescale}

In Section \ref{sec:results}, we postulated that $M_{\rm iso}$ will be larger if $t_{\rm peb}$ is shorter: cores may grow faster than they can create pressure bumps. It is unlikely that $t_{\rm peb}$ is shorter than our assumed $10^4$ yrs. The maximum rate of pebble accretion is given by $\sim$ $R_{\rm Hill}^2\Omega$ (i.e., cores gather all particles that enter their Hill spheres) and the associated accretion timescale is given by $\sim$ $10^4 (M/10\,M_\oplus)^{1/3}(r/{\rm 5 au})$ yrs \citep[see][their equation 44]{Lambrechts12}. Such rapid accretion can be maintained for particles with Stokes number $\sim$1 that settle to the midplane so that the solid disk scale height is smaller than the core's Hill radius, as would be the case for a nearly laminar disk. Future studies should explore how more accurate prescription of pebble accretion \citep[e.g.,][]{Ormel10} changes the preferred formation locations of gas giants.

\subsection{Disk Structure and Viscosity}

In this work, we used an inviscid, minimum mass solar nebula (MMSN). The MMSN is special in that its density profile ($\Sigma \propto r^{-1.5}$) produces a flat vortensity profile; there is no corotation torque so an embedded planet migrates according to the net Lindblad torque.
For any other disk with non-flat vortensity profiles, the corotation torque can either enhance, slow down, or even reverse the sign of the migration, depending on the disk profile, thermal structure, and the planet mass \citep{Paardekooper06, Baruteau08}.
Viscosity plays an important role for the corotation torque. In inviscid disks like the ones we considered, a dynamical corotation torque arises from the difference between the vortensity in the horseshoe region and the background disk \citep{Paardekooper14}. Under its influence, a shallower disk profile will reduce the migration rate, for example.
The degree to which the dynamical corotation torque may shift $r_{\rm div}$ remains to be investigated.

On the other hand, the classical, non-dynamical corotation torque favors viscous environments.
A sufficiently large viscosity can sustain a
corotation torque on the planet by re-establishing a vortensity gradient which would otherwise be erased by the libration of gas inside the horseshoe region \citep[][]{Maaset01,Masset10,Paardekooper11}.
\citet{Masset10} show that viscous diffusion sustains the corotation torque when
\begin{equation}
\label{eq:vis_sat1}
    \frac{r \nu}{\Omega x_{\rm s}^3} \gtrsim 0.1 \, ,
\end{equation}
where $\nu$ is the kinematic viscosity and $x_{\rm s}$ is the horseshoe half-width of the planet's co-orbital region. Given $\nu=\alpha h^2 \Omega$ and $x_{\rm s} \sim r\sqrt{(\Mp/M_\ast)/(h/r)}$ \citep{Fung15,Masset16}, \eqnref{eq:vis_sat1} translates to
\begin{equation}
\label{eq:vis_sat2}
    \alpha \gtrsim 2\times 10^{-3} \left(\frac{\Mp}{10 M_{\oplus}}\right)^{3/2}\left(\frac{h_{\rm p}/\rp}{0.035}\right)^{-7/2} \, .
\end{equation}

Another way viscosity affects migration is by eroding away the features of disk feedback so that planets need to be more massive than $M_{\rm fb}$ to halt.
We estimate that the gas pile-up in front of the planet and the deficit behind (see for example \figref{fig:den_evo}) have a size scale of about $h$, and so the viscous diffusion time for these features is
$t_{\rm vis} \sim hr/\nu = (\alpha \Omega h/r)^{-1}$.\footnote{The cores we consider take more than $\sim$ $10^4$ orbits to build up order-unity perturbations in inviscid disks; the perturbation is likely very weak in viscous disks. Assuming the degree of perturbation $\delta\Sigma/\Sigma < h/r$, the radial viscous flow speed across the gas pile-up/deficit is $\sim \nu /r$ rather than $\sim \nu/h$. In other words, the viscous flow does not ``see'' the density perturbations. The viscous diffusion timescale is then $t_{\rm vis} \sim hr/\nu$.}
By comparing $t_{\rm vis}$ to the time to form these features, or $t_{\rm delay}$ (\eqnref{eq:t_delay_emp}), we find that feedback becomes ineffective when:
\begin{align}
    \nonumber
    \alpha &\gtrsim 4 \times 10^{-4} \left(\frac{\Mp}{10 M_{\oplus}}\right)^{14/5}\left(\frac{h_{\rm p}/\rp}{0.035}\right)^{-42/5} \\
    &\sim 2 \times 10^{-4} \left(\frac{\Mp}{M_{\rm thermal}}\right)^{14/5} \, ,
    \label{eq:vis_fb}
\end{align}
While this is an order-of-magnitude estimate, we are encouraged that it agrees
well with the findings of \citet{Li09}; see their Figure 1. It should be noted that $t_{\rm delay}$ and therefore \eqnref{eq:vis_fb} are dependent on the mode of migration --- it would be different if migration is not predominantly driven by Lindblad torques. Future work accounting for corotation torques should provide a more complete picture of migration feedback.

Finally, in viscous disks, we may expect Type II migration to operate after the planets have opened gaps. In this case, some of the cold Jupiters in our models can migrate inward to become warm Jupiters. The reality of Type II migration is currently uncertain, as some recent work has found migration rates different from the classical viscous rate \citep{Duffell14,Durmann15}. Moreover, viscosity will reduce gap depth \citep{Fung14}, and partial gaps may facilitate Type III migration \citep{Papaloizou07}, further complicating the story.

Given these estimates and concerns, the results presented in this paper are most applicable when $\alpha$ is of order $10^{-4}$ or lower. Viscosities in real disks may be similar. Observations that probe the turbulent viscosity in protoplanetary disks such as HL tau \citep{Pinte16} and HD 163296 \citep{Flaherty15,Flaherty17} have found $\alpha$ on the level of $\sim$ $10^{-4}$--$10^{-3}$. These measurements are made at tens of au; near 1 au, where instabilities such as the vertical shear instability and magneto-rotational instability are both inoperative near the midplane \citep[e.g.,][]{Lin15,Gressel15,Bai17}, $\alpha$ is expected to be even lower.

\subsection{Warm Jupiters and the Lack of Wide-Orbit Gas Giants}

Both transit \citep{Dong13} and radial velocity \citep{Santerne16} surveys find a gradually declining population of gas giants at $\sim$1 au to $\sim$0.1 au. How can we explain the population of these ``warm'' Jupiters in the context of planet formation in inviscid (or nearly laminar) disks?

We find that gas giants are more likely to form closer to the star if they are born as dust-free worlds in more gas-heavy disks that live longer (see Section \ref{sec:final_loc}). Disk lifetime is not expected to play a significant role: observationally inferred gas disk lifetime ranges 1--10 Myrs \citep[e.g.,][]{mamajek09,alexander14} which changes $r_{\rm div}$ by only $\sim$50\% (see \eqnref{eq:rdiv}). In gas-heavy disks, $r_{\rm div}$ moves in but not by much: $r_{\rm div} \propto \Sigma_{1{\rm au}}^{-0.4}$ so to shorten $r_{\rm div}$ by factors of $\sim$10, $\Sigma$ needs to increase by factors of $\sim$300; such massive disks are susceptible to gravitational instability. It may be that dust-free gas accretion is a requirement for the formation of warm Jupiters. This is not necessarily in contradiction with \citet{Thorngren16} who report significant heavy element enrichment in warm Jupiters. Cores that nucleate into warm Jupiters can build dust-free envelope prior to the runaway gas accretion; subsequent pollution by drifting solids and/or the erosion of the core can enhance the overall contents of heavy elements in the envelopes of warm Jupiters.

Alternatively, changes in disk structure and viscosity may alter migration rates and place Jupiters closer to the star, as discussed in the previous section. It is also possible that collisions between multiple small cores to larger cores in the inner disk can seed the formation of warm Jupiters. To birth gas giants, core-core collisions must occur in gas-rich environments. Gas dynamical friction will render these early giant impacts low-probability events which may explain the rarity of warm Jupiters. Quantifying the rate of warm Jupiter formation by giant impacts in gas-rich nebulae is the subject of our ongoing work.

Even though the outer regions of protoplanetary disks are the likely breeding grounds for gas giants, gas giants may still be rare at large distances. In fact, statistical analyses of directly imaged planets find massive gas giants (5--13 $M_{\rm Jup}$) to occur only around $\sim$0.6\% of stars at distances of 30--300 au (\citealt{bowler16};
see also \citealt{Meshkat17} who quote an occurrence rate of $\sim$0.7\% at distances of 10--1000 au). Combining radial velocity trends with imaging, \citet{Bryan16} report that the occurrence rate of gas giant companions to RV-detected systems tends to fall off at distances beyond $\sim$10 au. Whether a given core nucleates into gas giants or not depends largely on how massive the core is \citep[e.g.,][]{Ikoma00,Rafikov06,Piso14,Lee14}. In other words, the core growth timescale and the availability of solids in the disk are the likely determining factors of the planet population that emerges in a given disk. In the present paper, we have assumed solids to be infinitely replenished; in future work, we will relax this assumption and investigate in detail how the global radial drift of pebbles shapes the overall exoplanet demographics at large distances.


\acknowledgements
We thank Eugene Chiang, Ruobing Dong, Paul Duffell, Anders Johansen, Heather Knutson, Michiel Lambrechts, Renu Malhotra, and Chris Ormel for helpful discussions.
An anonymous referee provided an encouraging and helpful report.
This work was performed under contract with the Jet Propulsion
Laboratory (JPL) funded by NASA through the Sagan Fellowship Program
executed by the NASA Exoplanet Science Institute. EJL acknowledges support from a Sherman Fairchild Fellowship at Caltech.

\bibliographystyle{aasjournal}
\bibliography{Lit}

\end{document}